\documentclass[preprint2,times,tighten,twocolappendix,trackchanges]{aastex6}
\usepackage{amsmath,amstext}
\usepackage[figure,figure*]{hypcap}
\usepackage{subfigure,graphicx}
\usepackage{newtxmath} 
\usepackage{lmodern}
\usepackage[T1]{fontenc}

\newcommand{\Msun}{\ifmmode {M_{\odot}}\else${M_{\odot}}$\fi}
\newcommand{\Rsun}{\ifmmode {R_{\odot}}\else${R_{\odot}}$\fi}
\newcommand{\Lsun}{\ifmmode {L_{\odot}}\else${L_{\odot}}$\fi}
\newcommand{\lapprox }{{\lower0.8ex\hbox{$\buildrel <\over\sim$}}}
\newcommand{\gapprox }{{\lower0.8ex\hbox{$\buildrel >\over\sim$}}}
\def\amin{\ifmmode^{\prime}\else$^{\prime}$\fi}
\def\asec{\ifmmode^{\prime\prime}\else$^{\prime\prime}$\fi}

\newcommand{\degree}{\ifmmode {^\circ}\else$^\circ$\fi}

\newcommand{\Ro}{\ifmmode {R_o}\else$R_o\ $\fi}
\newcommand{\lha}{\ifmmode {L_{H\alpha}/L_{bol}}\else$L_{H\alpha}/L_{bol}$\ \fi}

\slugcomment{DRAFT \today}
\shorttitle{{\it K2} Periods for Praesepe}
\shortauthors{Douglas et al.}

\begin{document}

\title{Poking the Beehive from Space: K2 Rotation Periods for Praesepe}
\author{S.~T.~Douglas\altaffilmark{1},
M.~A.~Ag{\" u}eros\altaffilmark{1},
K.~R.~Covey\altaffilmark{2},
A.~Kraus\altaffilmark{3}
}
\altaffiltext{1}{Department of Astronomy, Columbia University, 550 West 120th Street, New York, NY 10027, USA}
\altaffiltext{2}{Department of Physics and Astronomy, Western Washington University, Bellingham WA 98225, USA}
 \altaffiltext{3}{Department of Astronomy, University of Texas at Austin, 2515 Speedway, Stop C1400, Austin, TX 78712, USA}

\begin{abstract}
We analyze {\it K2} light curves for 794 low-mass ($1\ \gapprox\ M_*\  \gapprox\ 0.1$~\Msun) members of the $\approx$650-Myr-old open cluster Praesepe, and measure rotation periods ($P_{rot}$) for 677 of these stars. 
We find that half of the rapidly rotating $\gapprox$0.3~\Msun\ stars are confirmed or candidate binary systems. 
The remaining $\gapprox0.3$~\Msun\ fast rotators have not been searched for companions, and are therefore not confirmed single stars.  
We found previously that nearly all rapidly rotating $\gapprox$0.3~\Msun\ stars in the Hyades are binaries, but we require deeper binary searches in Praesepe to confirm whether binaries in these two co-eval clusters have different $P_{rot}$ distributions.
We also compare the observed $P_{rot}$ distribution in Praesepe to that predicted by models of angular-momentum evolution. 
We do not observe the clear bimodal $P_{rot}$ distribution predicted by \citet{brown2014} for $>$0.5~\Msun\ stars at the age of Praesepe, but 0.25$-$0.5~\Msun\ stars do show stronger bimodality.
In addition, we find that $>$60\% of early M dwarfs in Praesepe rotate more slowly than predicted at 650 Myr by \citet{matt2015}, which suggests an increase in braking efficiency for these stars relative to solar-type stars and fully convective stars.
The incompleteness of surveys for binaries in open clusters likely impacts our comparison with these models, since the models only attempt to describe the evolution of isolated single stars. 
\end{abstract}

\maketitle

\section{Introduction}
In examining the evolution of angular momentum and activity in late-type stars, the Hyades and Praesepe ($\alpha =$ 04:27, $\delta = +15$:52 and $\alpha =$ 08:40:24, $\delta = +$19:41, respectively),  two $\approx$650-Myr-old open clusters, form a crucial bridge between young open clusters \citep[such as the Pleiades, at $\approx$125 Myr; e.g.,][]{Covey2016,rebull2016-1} and older field dwarfs \citep[$\geq$2 Gyr; e.g.,][]{kiraga2007}. This paper is the fourth in our study of these linchpin clusters.

In \citet[][hereafter Paper I]{agueros11}, we presented new rotation periods ($P_{rot}$) for 40 late-K to mid-M Praesepe members measured from Palomar Transient Factory \citep[PTF;][]{nick2009,rau2009} data. 
We also tested models of angular-momentum evolution, which describe the evolution of stellar $P_{rot}$ as a function of color and mass. 
We used the semi-empirical relations of \citet{barneskim2010} and \citet{barnes2010} to evolve the sample of Praesepe periods. 
Comparing the resulting predictions to periods measured in M35 and NGC 2516 ($\approx$150 Myr) and for kinematically selected young and old field star populations (1.5 and 8.5 Gyr, respectively), we found that stellar spin-down may progress more slowly than described by these relations.

In \citet[][hereafter Paper II]{douglas2014}, we extended our analysis to the Hyades, combining new $P_{rot}$ measured with All Sky Automated Survey \citep[][]{ASAS} data (Cargile et al.\ in prep) with those obtained by \citet{radick1987,radick1995}, \citet{scholz2007}, \citet{scholz2011}, and \citet{delorme2011}. We combined these data with new and archival optical spectra to show that the transition between 
magnetically inactive and active stars happens at the same mass in both clusters, as does the transition from a partially active population to one where every star is active. 
Furthermore, we determined that Praesepe and the Hyades are following identical rotation-activity relations, and that the mass-period relation for the combined clusters transitions from an approximately single-valued sequence to a wide spread in $P_{rot}$ at a mass $M_* \approx$~0.6$-$0.7~\Msun, or a spectral type SpT $\approx$ M0.

In \citet[][hereafter Paper III]{douglas2016}, however, 
after adding $P_{rot}$ from \citet{prosser1995}, \citet{hartman2011}, and our observations with the re-purposed {\it Kepler} mission  \citep[{\it K2};][]{howell2014},
and after removing all confirmed and candidate binaries from the Hyades's mass-period plane, 
we found that nearly all \textit{single} Hyads with $M_*\ \gapprox\ 0.3\ \Msun$ are slowly rotating. 
We also found that the more recent, theoretical models for rotational evolution of \citet{reiners2012} and \citet{matt2015} predict faster rotation than is actually observed at $\approx$650~Myr for $\lapprox$0.9 \Msun\ stars. 
The dearth of single $\gapprox$0.3~\Msun\ rapid rotators 
indicates that magnetic braking is more efficient than previously thought, and that age-rotation studies must account for multiplicity.

We now present $P_{rot}$ measurements for 677 Praesepe members measured from {\it K2} data. 
We describe the membership catalog and archival $P_{rot}$ we used in Section~\ref{data}, and our {\it K2} light curves and period-measuring algorithm in Section~\ref{prot}. 
To examine the impact of multiplicity on the mass-period plane, we attempt to identify binaries in Praesepe; we  discuss these efforts in Section~\ref{binaries}.
We present our results, including their potential implications for calibrating angular momentum evolution, in Section~\ref{res}. 
We conclude in Section~\ref{concl}.

\section{Existing Data}\label{data}
\subsection{Cluster Catalog}\label{cats}
We continue to use the Praesepe membership catalog presented in Paper II, which includes 1130 cluster members with membership probabilities  $P_{mem}\geq50\%$ as calculated by \citet{adam2007} and 39 previously identified members too bright to be included by those authors in their catalog for the cluster. 
We assign these bright stars $P_{mem}=100$\%.
We also continue to use the photometry and stellar masses presented in table 5 of Paper II. 
For most of our analysis, as in that work, we include only the 1099 stars with $P_{mem}\geq70\%$. 

\begin{figure}[t]
\centerline{\includegraphics[width=\columnwidth]{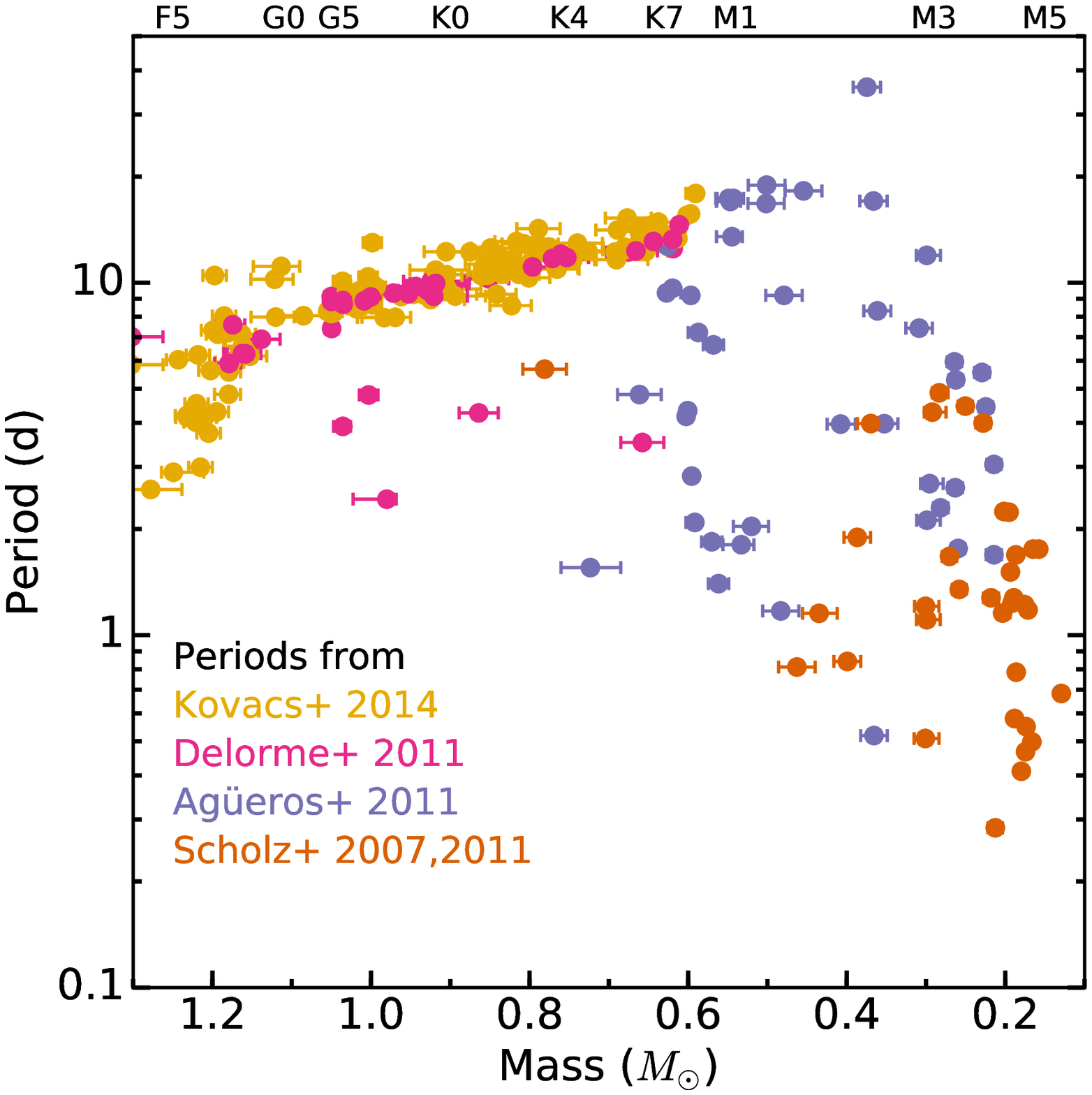}}
\caption{Praesepe mass-period plane  showing all literature $P_{rot}$ before the addition of the new {\it K2} data. Periods for F, G, and K dwarfs were measured from all-sky surveys by \citet[][yellow]{kovacs2014} and \citet[][pink]{delorme2011}, while periods for M dwarfs were measured from targeted surveys of the cluster by \citet[][purple]{agueros11}, \citet[][]{scholz2007}, and \citet[][both orange]{scholz2011}. Approximate spectral types corresponding to the plotted masses are indicated at the top of the plot.}
\label{fig:periodmass_old}
\end{figure}

\subsection{Archival Rotation Periods}\label{archprot}

In Papers I and II, we combined $P_{rot}$ measurements from PTF data with $P_{rot}$ measurements from \citet{scholz2007}, \citet{scholz2011}, and \citet{delorme2011} to produce a catalog of 135 known rotators in Praesepe.\footnote{For details on these data, see Paper II and the original papers.} 
Eighty-three of these stars have a \citet{adam2007} $P_{mem} > 95\%$.

To this catalog we now add 180 $P_{rot}$ measurements from \citet{kovacs2014}; 174 of these stars have $P_{mem}\geq70\%$. 
Forty-four stars have previous $P_{rot}$ measurements by other authors: 
the majority of these measurements are consistent to within 0.5 d, but 13 stars have significantly discrepant $P_{rot}$ measurements (see Table~\ref{tab:protdiff}). 
In all 13 cases, \citet{kovacs2014} measure the $P_{rot}$ to be at least twice as long as previous authors. This discrepancy undermines the validity of the other \citet{kovacs2014} $P_{rot}$ values, and we therefore retain the previous literature $P_{rot}$ wherever possible.

In total, we add 136 rotators with non-\textit{K2} $P_{rot}$ to our Praesepe catalog, including 131 with $P_{mem} > 70\%$. 
The mass-period data for Praesepe members with existing $P_{rot}$ measurements is shown in Figure \ref{fig:periodmass_old}. 

\floattable
\begin{deluxetable}{llccccc}
[t]\tablewidth{0pt}\tabletypesize{\scriptsize}
\tablecaption{Praesepe members with discrepant $P_{rot}$ measurements \label{tab:protdiff}}
\tablehead{\colhead{Name\tablenotemark{a}} & \colhead{EPIC} & \colhead{Ag\"ueros et al.\ 2011} & \colhead{Delorme et al.\ 2011} & \colhead{Scholz et al.\ 2007, 2011} & \colhead{Kovacs et al.\ 2014} & \colhead{{\it K2}}\\ \colhead{ } & \colhead{ } & \colhead{$P_{rot}$ (d)} & \colhead{$P_{rot}$ (d)} & \colhead{$P_{rot}$ (d)} & \colhead{$P_{rot}$ (d)} & \colhead{$P_{rot}$ (d)}}
\startdata
\nodata & 211885995 & 9.20 & \nodata & \nodata & 18.13 & 9.16 \\
AD 1508 & 212009427 & 1.55 & \nodata & \nodata & 11.22 & 1.56 \\
AD 1512 & \nodata & 9.64 & \nodata & \nodata & 19.15 & \nodata \\
AD 2182 & 211734093 & \nodata & \nodata & \nodata & 15.87 & 18.22 \\
AD 2509 & 211970613 & \nodata & \nodata & 0.50 & \nodata & 1.01 \\
AD 2527 & 211939989 & \nodata & \nodata & 0.47 & \nodata & 0.92 \\
AD 2552 & 211989299 & 25.36 & \nodata & \nodata & \nodata & 12.84 \\
AD 2802 & 211980450 & \nodata & \nodata & 0.51 & \nodata & 1.02 \\
AD 3128 & \nodata & \nodata & 3.52 & \nodata & 14.17 & \nodata \\
AD 3663 & 211773459 & \nodata & \nodata & \nodata & 17.91 & 5.94 \\
HSHJ 15 & 211971354 & 9.36 & \nodata & \nodata & 17.46 & 8.26 \\
HSHJ229 & 211938988 & \nodata & \nodata & 2.29 & \nodata & 1.09 \\
HSHJ421 & 211944193 & \nodata & \nodata & 0.28 & \nodata & 0.48 \\
HSHJ436 & 211988700 & \nodata & \nodata & 4.87 & \nodata & 6.46 \\
JS140 & 211930699 & \nodata & \nodata & \nodata & 13.35 & 6.74 \\
JS298 & 211945362 & \nodata & \nodata & 4.29 & \nodata & 9.16 \\
JS313 & 211992053 & \nodata & \nodata & 5.76 & \nodata & 5.08 \\
JS379 & 212013132 & \nodata & 4.27 & \nodata & 12.78 & 2.13 \\
JS418 & 211954582 & \nodata & \nodata & 3.27 & 12.75 & 3.19 \\
JS432 & \nodata & 2.09 & \nodata & \nodata & 8.36 & \nodata \\
JS503 & 212019252 & \nodata & 9.95 & \nodata & \nodata & 11.26 \\
JS547 & 211923502 & \nodata & \nodata & \nodata & 10.73 & 12.07 \\
JS655 & 211896596 & \nodata & \nodata & \nodata & 5.85 & 2.97 \\
JS719 & 211989620 & \nodata & \nodata & 1.21 & \nodata & 0.88 \\
KW 30 & 211995288 & \nodata & 3.91 & \nodata & 7.97 & 7.80 \\
KW141 & 211940093 & \nodata & 9.42 & \nodata & 9.79 & 4.89 \\
KW172 & 211975426 & \nodata & \nodata & \nodata & 12.22 & 6.26 \\
KW256 & 211920022 & \nodata & 4.80 & \nodata & 9.76 & 4.67 \\
KW267 & 211970147 & \nodata & \nodata & 5.68 & 11.89 & 11.60 \\
KW301 & 211936906 & \nodata & \nodata & \nodata & 7.58 & 8.76 \\
KW304 & 211996831 & \nodata & \nodata & \nodata & 8.79 & 4.37 \\
KW336 & 211911846 & \nodata & 8.89 & \nodata & 9.12 & 4.30 \\
KW367 & 211975006 & \nodata & \nodata & \nodata & 6.04 & 3.07 \\
KW401 & 211909748 & \nodata & 2.43 & \nodata & 9.61 & 2.42 \\
KW434 & 211935518 & \nodata & \nodata & \nodata & 8.27 & 4.18 \\
KW533 & 211954532 & \nodata & \nodata & \nodata & 8.29 & 9.27 \\
KW563 & 211970427 & 4.33 & \nodata & 4.85 & \nodata & 4.38 \\
KW566 & 211988628 & \nodata & \nodata & \nodata & 15.25 & 7.95 \\
KW570 & 211983725 & 4.18 & \nodata & 4.27 & 16.81 & 4.22 \\
\enddata
\tablecomments{Only cluster members with at least two $P_{rot}$  measurements that differ by at least 10\% are shown here. An additional 215 cluster members have at least two $P_{rot}$ measurements that agree to within 10\%.}
\tablenotetext{a}{Literature name given in \citet{adam2007}. All are standard SIMBAD identifiers, except AD\#\#\#\#, which correspond to stars in \citet{adams2002}. 
}
\end{deluxetable}

\section{Measuring Rotation Periods with K2}\label{prot}

\begin{figure}[t]
\centerline{\includegraphics[width=\columnwidth]{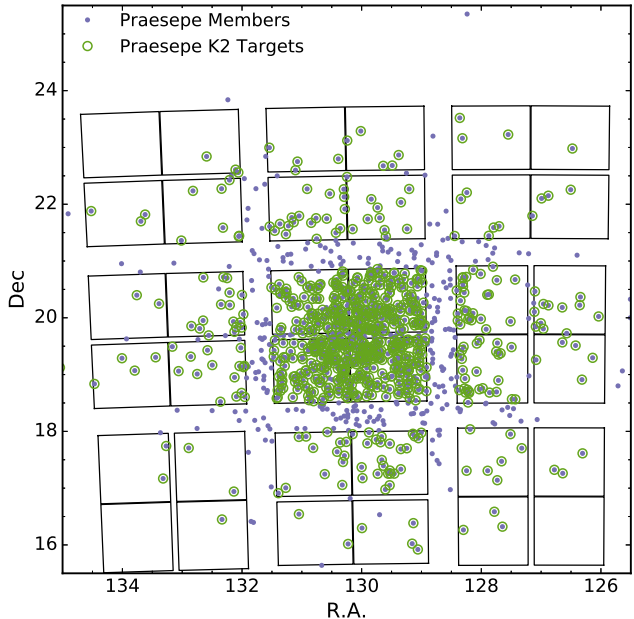}}
\caption{All Praesepe members (blue dots) and 794 K2 targets (green circles), with the {\it K2} chip edges overlaid. 
The entire cluster fits within the {\it K2} field-of-view, although many members still fall in the chip gaps. 
Two of the detector modules were no longer functioning by the time Campaign 5 started, but no Praesepe members fell on these modules. 
}
\label{fig:k2fov}
\end{figure}

{\it K2} targeted Praesepe in its Campaign 5. We analyze the resulting long-cadence data for 794 Praesepe members identified in Section~\ref{cats} and with {\it Kepler} magnitudes $K_p>9$ mag and masses $M_*<1.5$~\Msun. These limits exclude saturated stars as well as stars with radiative outer layers, which are outside of the scope of this work.
The distribution of targets on the {\it K2} imager is shown in Figure~\ref{fig:k2fov}.  Of the 794 targets, 749 have a  \citet{adam2007} $P_{mem}>70\%$.

\subsection{K2 Light Curves}\label{k2lc}
The pointing in {\it K2} is held in an unstable equilibrium against solar pressure by the two functioning reaction wheels. The spacecraft rolls about the boresight by up to 1 pixel at the edge of the focal plane.
To correct for this, thrusters can be fired every 6 hr (if needed) to return the spacecraft to its original position.
This drift causes stars to move in arcs on the focal plane, inducing a sawtooth-like signal in the 75-d light curve for each star \citep{vancleve2016}.

Several groups have developed methods for extracting photometry and removing the effect of the pointing drift from the raw light curve. 
We tested the light curves produced using several detrending methods \citep[][]{vanderburg2014,aigrain2016,luger2016}, as well as our own (Paper III). 
We chose to use the light curves generated by the {\it K2} Systematics Correction method \citep[K2SC;][]{aigrain2016} for our analysis, 
as this approach does the best job of removing systematics and long-period trends, which can bias period measurements or completely wash out periodic signals.

\citet{aigrain2016} use a semi-parametric Gaussian process model to correct for the spacecraft motion. 
These authors begin with the light curves and centroid positions produced by the {\it Kepler} Science Operations Center pipeline. 
They then simultaneously model the position-dependent, time-dependent, and white-noise components of the light curve. 
The time-dependent component should describe the intrinsic variability of the star, and the position-dependent component should describe the instrumental signal resulting from the spacecraft roll. 
In cases where a significant period between 0.05 and 20~d is detected in the raw light curve, \citet{aigrain2016} use a quasi-periodic kernel to describe the time-dependent trend; otherwise these authors use a squared-exponential kernel.

Since we wish to measure stellar variability, we remove only the position-dependent trend. 
The provided light-curve files include the position-dependent, time-dependent, and white-noise components in separate columns for both the simple aperture photometry (SAP) and pre-search data conditioning (PDC)
pipeline light curves \citep{vancleve2016}.
We use the PDC light curves, and compute the final light curve for our analysis by adding the white noise and time-dependent components, and then subtracting the median of the time-dependent component.\footnote{\url{https://archive.stsci.edu/missions/hlsp/k2sc/hlsp_k2sc_k2_llc_all_kepler_v1_readme.txt}} 

\subsection{Measuring Rotation Periods}\label{prot_meas}
We use the \citet{press1989} FFT-based Lomb-Scargle algorithm\footnote{Implemented as {\it lomb\_scargle\_fast} in the {\it gatspy} package; see \url{https://github.com/astroML/gatspy}.} to measure rotation periods. We compute the Lomb-Scargle periodogram power for 3$\times$10$^4$ periods ranging from 0.1 to 70~d (approximately the length of the Campaign).

The periodogram power, which is normalized so that the maximum possible power is 1.0, is the first measurement of detection quality. The normalized power, $P_{LS}$, is related to the ratio of $\chi^2$ for the sinusoidal model to $\chi^2_0$ for a pure noise model \citep{ivezic2013}:
\begin{equation}
P_{LS} = 1-\frac{\chi^2}{\chi^2_0}.
\end{equation}

A higher $P_{LS}$ indicates that the signal is more likely sinusoidal, and a lower $P_{LS}$ indicates that it is more likely noise.
Therefore, $P_{LS}$ gives some information about the relative contributions of noise and periodic modulation to the light curve.
We do not impose a global minimum value for $P_{LS}$. Instead, we compute a minimum significance threshold for each light curve. 

We identify periodogram peaks using the \textit{scipy.signal.argrelextrema} function, and define a peak as any point in the periodogram higher than at least 100 of the neighboring points. 
This value was chosen after some trial and error, and has the benefit of automatically rejecting most long period trends, because the periodogram is logarithmically sampled and has fewer points at long periods. 
Long period trends appear as a peak near 60--70 d with a series of harmonic peaks; these are generally rejected by \textit{argrelextrema}. 
When there is a sinusoidal stellar signal in the light curve, it dominates the periodogram above any trends and is detected by \textit{argrelextrema}.

We determine minimum significance thresholds for the periodogram peaks using bootstrap re-sampling, as in Paper III. 
We hold the observation epochs fixed and randomly redraw and replace the flux values to produce new scrambled light curves. 
We then compute a Lomb-Scargle periodogram for the scrambled light curve, and record the maximum periodogram power. 
We repeat this process 1000 times, and take the 99.9$^{\rm th}$ percentile of peak powers as our minimum significance threshold for that light curve. 
A peak in our original light curve is significant if its power is higher than this minimum threshold, which is listed in Table~\ref{tab:k2}. 
We take the highest significant peak as our default $P_{rot}$ value; twenty-three of our targets show no significant periodogram peaks.

\begin{figure}[t]
   \centering
   \subfigure[Neither detected period matches the observable repeats in the full light curve; this may be a case of rapid spot evolution or differential rotation. We set Q~$=2$ as we cannot determine the correct period.]{
   \includegraphics[width=0.95\columnwidth]{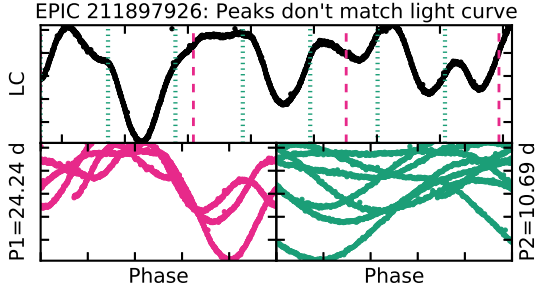}
   }
   \subfigure[Double-dip structure, periodogram selects half of the likely true period. We select the longer period and set Q~$=0$.]{
   \includegraphics[width=0.95\columnwidth]{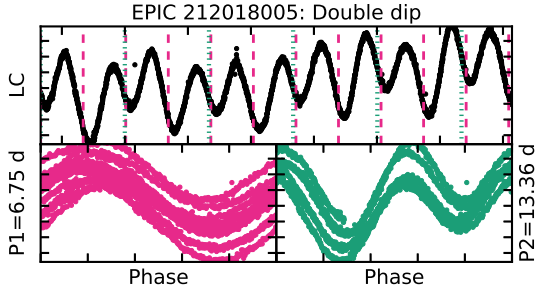}
   }
   \subfigure[A non-repeating trend is detected with high periodogram power; we set Q~$=2$.]{
   \includegraphics[width=0.95\columnwidth]{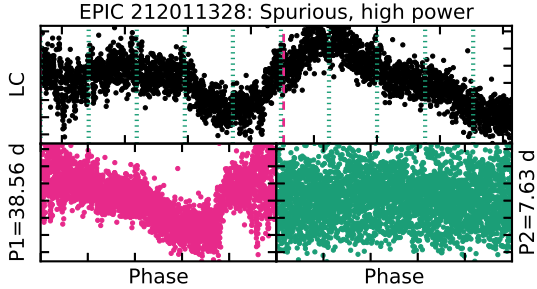}
   }
   \subfigure[There are two clear periods in the light curve. We set Q~$=0$, and we flag this target as definitely multiperiodic and therefore a candidate binary.]{
   \includegraphics[width=0.95\columnwidth]{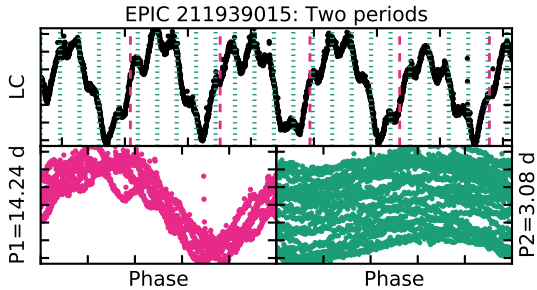}
   }
\caption{Examples of the light curve effects discussed in Section \ref{validation}. 
Vertical lines at intervals of the detected period are overlaid on each full light curve, as in Figure~\ref{fig:k2lc}. 
The phase-folded light curves corresponding to the first and second highest periodogram peaks are also shown. 
}\label{fig:pbs}
\vspace{0.6cm}
\end{figure}

\subsection{Validating the Measured Rotation Periods}\label{validation}

We combine automated and by-eye quality checks to validate the $P_{rot}$. 
The automated check comes from the peak periodogram power along with the number of, and power in, periodogram peaks beyond the first. 
Following \citet{Covey2016}, we label a periodogram as clean if there are no peaks with more than 60\% of the primary peak's power. The presence of such peaks  may indicate that the $P_{rot}$ measurement is incorrect. 
The clean flag is included in Table \ref{tab:k2}; only 46 {\it K2} detections are not clean.

\begin{figure}[t]
\centerline{\includegraphics[width=\columnwidth]{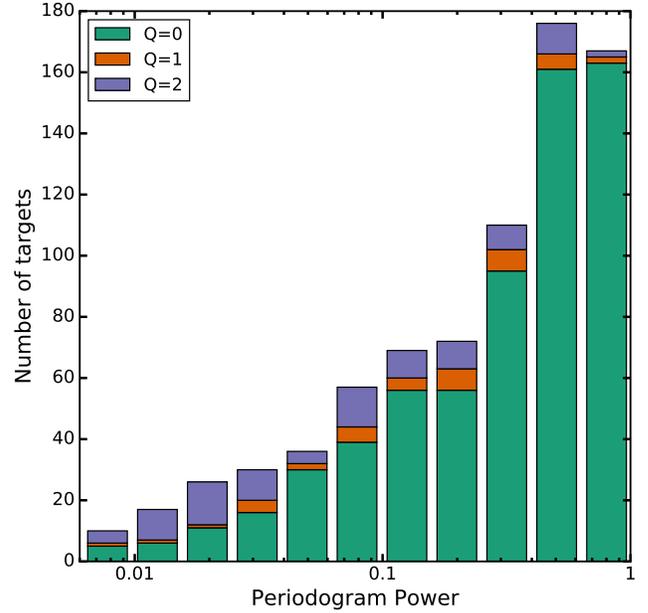}}
\caption{Histogram of periodogram powers from our sample; colors correspond to the flags assigned during our visual inspection of all $P_{rot}$ detections. Spurious detections (Q~$=2$, purple) occur at a low rate across the full range of periodogram powers, motivating our visual inspection. }
\label{fig:powers}
\end{figure}

In addition, since instrumental signals can occasionally be detected at high significance, we inspect the periodograms and phase-folded light curves by eye to confirm detections.
Clearly spurious detections are flagged as Q~$=2$, and questionable detections as Q~$=1$.
This is similar to the approach used in Paper III, but we are more generous here and try to identify only the most obvious bad detections.
In total, we remove 94 light curves.
Additionally, a Q~$=3$ flag indicates that there were no significant periodogram peaks; as noted earlier, this occurred for 23 stars. 

The Q flag is separate from the clean/not-clean classification, and we do not change the Q value based on the clean/not-clean classification. 
We consider $P_{rot}$ measurements with a clean periodogram and Q~$=0$ to be high-quality detections. 
In cases where we measure a \textit{K2} $P_{rot}$ for a star with a $P_{rot}$ in the literature, the agreement is generally excellent (see Section \ref{consistency}). This indicates that our methods produce reasonable and valid $P_{rot}$ measurements.

\begin{figure*}[p]
\centerline{\includegraphics[width=\textwidth]{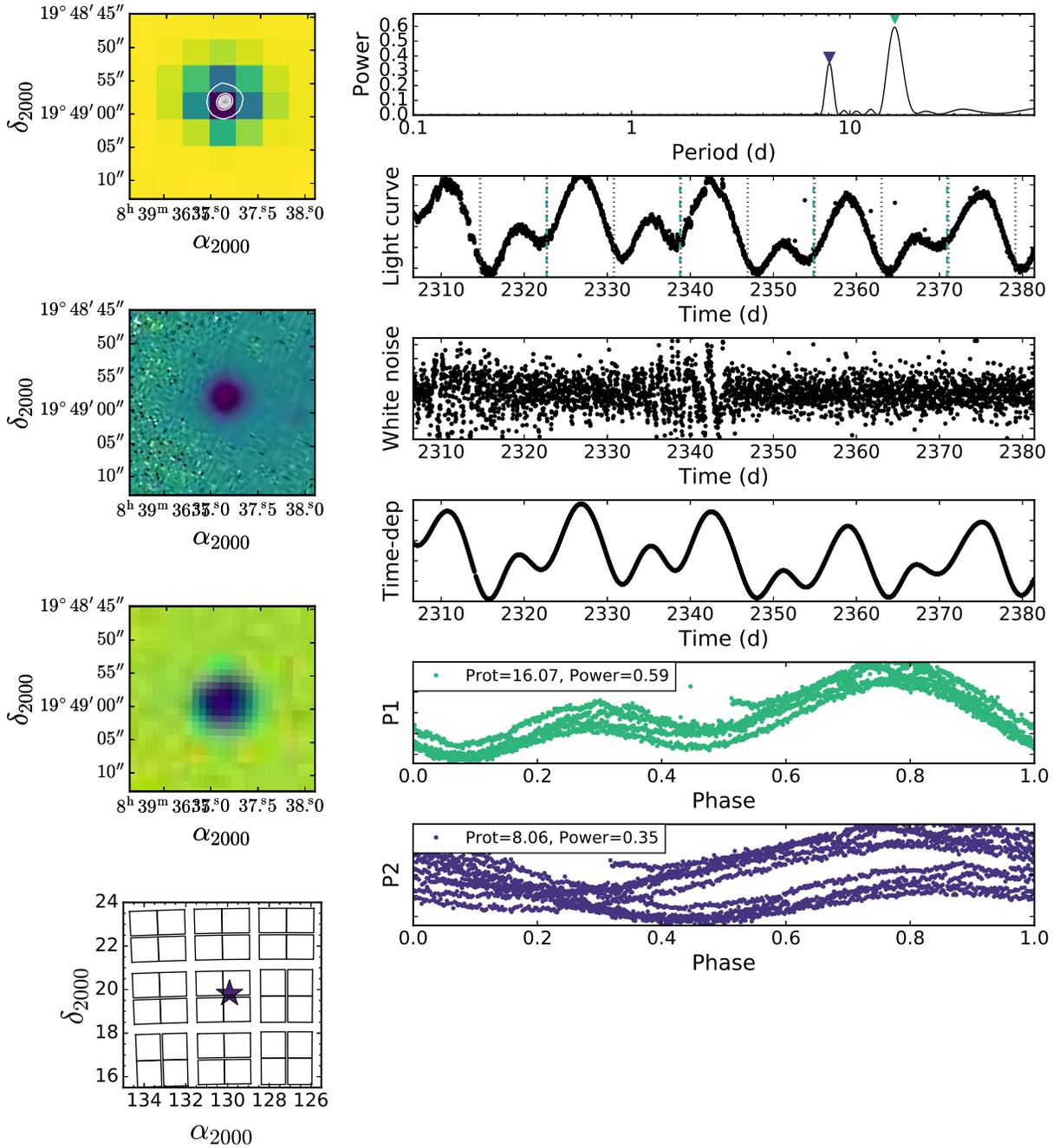}}
\caption{An example of the plots used to inspect period detections and check for neighboring stars. 
Left column, top to bottom: {\it K2} pixel stamp with SDSS $r$-band image overlaid as a contour; SDSS $r$-band image rotated into the {\it K2} frame; DSS red image rotated into the {\it K2} frame; and the target's position within the {\it K2} Campaign 5 field of view. 
Right column, top to bottom: Lomb-Scargle periodogram with (up to) the three highest significant peaks indicated by inverted triangles; the light curve corrected for spacecraft drift; the white-noise component of the light curve; the time-dependent component; and the light curve phase-folded on (up to) the three most significant periods. 
The colors of the markers indicating the peaks in the periodogram correspond to the colors of the phase-folded light curves. 
Versions of this plot for every {\it K2} target analyzed are available as an electronic figure set.}
\label{fig:k2lc}
\end{figure*}

Following \citet{mcquillan2013}, we plot the full light curve and, for $P_{rot} >2$ d, vertical dashed lines at intervals of the detected period. 
We check that light curve features repeat over several intervals.
We identify six cases where the phased light curve looks reasonable, but the pattern identified by eye does not match that detected in the periodogram (see Figure~\ref{fig:k2lc} and top panel of Figure~\ref{fig:pbs}), and we flag these with Q~$=2$.

\begin{deluxetable*}{llllll}
\tablecaption{Companions to Praesepe members with measured $P_{rot}$\label{tab:bin}}
\tablehead{
\colhead{} & \colhead{} & \colhead{} 
& \colhead{Binary} & \colhead{Triple} & \colhead{}\\
\colhead{Name\tablenotemark{a}} & \colhead{EPIC} & \colhead{2MASS} 
& \colhead{Type} & \colhead{Type} & \colhead{Source}
}
\startdata
KW350 & 211980142 & J08405693+1956055 &  SB2 & \nodata & \citet{dickens1968, patience2002} \\
JS401 & 211896450 & J08405866+1840303 &  Photometric & \nodata & \citet{douglas2014} \\
JS402 & \nodata & J08405968+1822045 & Photometric & \nodata & \citet{douglas2014} \\
KW365 & 211923188 & J08410737+1904165 &  SB1 & SB1 & \citet{bolte1991, mermilliod1994, mermilliod1999} \\ 
 & & & & & \citet{bouvier2001, patience2002, halbwachs2003} \\
 & & & & & \citet{mermilliod2009} \\
KW367 & 211975006 & J08410961+1951187 & SB1 & SB1 & \citet{mermilliod1994, mermilliod1999} \\
 & & & & & \citet{halbwachs2003, mermilliod2009,douglas2014} \\
KW371 & 211952381 & J08411002+1930322 &  Photometric & \nodata & \citet{mermilliod1999, patience2002} \\
KW368 & 211972627 & J08411031+1949071 & SB1 & \nodata & \citet{mermilliod1999, halbwachs2003} \\
 & & & & & \citet{mermilliod2009} \\
JS418 & 211954582 & J08411319+1932349 &  Photometric & \nodata & \citet{hodgkin1999,douglas2014} \\
KW375 & 211979345 & J08411377+1955191 & SB & \nodata & \citet{johnson1952} \\
KW385 & 211935741 & J08411840+1915395 & Visual & \nodata & \citet{patience2002,douglas2014} \\
\enddata 
\tablecomments{This table is available in its entirety in a machine-readable form in the online journal. A portion is shown here for guidance regarding its form and content.}
\tablenotetext{a}{Literature name given in \citet{adam2007}. All are standard SIMBAD identifiers, except AD\#\#\#\#, which correspond to stars in \citet{adams2002}.}
\end{deluxetable*}

We also identify 13 light curves where the dominant periodogram peak is likely for half of the true period and there is double-dip structure in the light curve  (see second panel, Figure~\ref{fig:pbs}). There is typically a periodogram peak at this longer period that is weaker than the dominant peak. This feature is common in stellar light curves and usually attributed to symmetrical spot configurations and/or an evolving spot pattern on the stellar surface \citep{mcquillan2013}.

In most Q~$=2$ cases, the phase-folded light curve does not look sinusoidal (third panel, Figure \ref{fig:pbs}), and the light curve is likely just noise. 
We also remove three stars where the saturation strip from a nearby star crosses the target pixel stamp, and one where the target is extended and likely a galaxy based on its Sloan Digital Sky Survey Data Release 12 \citep[SDSS DR12;][]{alam2015} image.

As part of our visual inspection, we also note cases where two or more periods are detected, i.e., due to multiple stars being present in the aperture (fourth panel, Figure \ref{fig:pbs}), or where there we find evidence for spot evolution (second panel,  Figure~\ref{fig:pbs}). We assign flags for targets with multiple periods and with spot evolution: ``Y'' for yes, ``M'' for maybe, and ``N'' for no.  

Finally, we note any other interesting light curve features, typically transits or eclipses (see Appendix~\ref{eclipses} for discussion of the latter light curves). An example set of our inspection plots is shown in Figure~\ref{fig:k2lc}, and the plots for all of our objects are available as an electronic figure set.

\subsection{Photometric Amplitudes}

We measure the amplitude of variability for a given star using the 10$^{\rm th}$ and 90$^{\rm th}$ percentiles ($P_{10}$ and $P_{90}$) of the light curve in counts.
We calculate the amplitude in magnitudes as
\begin{equation}\label{eq:amp}
2.5 \times \frac{\left[\textrm{log}_{10}\left(P_{90}\right) - \textrm{log}_{10}\left(P_{10}\right)\right]}{2}.
\end{equation}

This number may be slightly misleading, however, in cases where the median flux level varies over the course of the Campaign (a minor example is shown in the second panel of Figure \ref{fig:pbs}). Therefore, we also calculate a smoothed version of the phase-folded light curve, and measure the amplitude as the percent difference between the maximum and minimum values of the smoothed light curve. 
This method, already used in Paper III, tends to underpredict the amplitude of very fast rotators. 
We list both amplitudes in Table \ref{tab:k2}, but use the amplitude calculated using Equation \ref{eq:amp} for all analysis below. 
Our results do not change significantly when using the amplitudes calculated by either method.


\section{Binary Identification} \label{binaries}

We identify as many binary systems as possible among our {\it K2} targets, both to account for tidal effects and the more mundane impact of two (or more) stars blended on the chip.
We denote all confirmed and candidate binaries in our analysis below.

Binary companions may impact rotational evolution via gravitational or magnetic interactions. 
Stars in very close binaries can exert tidal forces on each other, spinning them up or down more rapidly than predicted for a single star \citep[e.g.,][]{meibom2005,zahn2008}. 
These systems are also close enough for one star to interact with the other's large-scale magnetic field. 
And at the earliest evolutionary stages, a companion may truncate the protoplanetary disk, minimizing the impact of magnetic braking and allowing the young star to spin faster than its single counterparts \citep[e.g.,][]{rebull2004,meibom2007,cieza2009}. 
Any of these effects could result in different angular-momentum evolution paths for stars with and without companions.

Furthermore, binaries may contaminate our analysis of $P_{rot}$ distributions. 
If two stars are blended in ground-based images as well, the additional flux from the companion may cause us to overestimate $L_{bol}$ and $M_*$. 
A companion may also dilute the rotational signal, leading to underestimated photometric amplitudes or masking the rotation of the fainter component altogether. 
In the case of two detected periods, it is impossible to tell which signal comes from which star.
These effects can cause stars to be misplaced in the mass-period plane, leading us to misidentify trends or transition periods.

\subsection{Visual Identification}\label{visbin}
We examine a co-added \textit{K2} image, a Digital Sky Survey (DSS) red image, and an SDSS \citep{alam2015} $r$-band image of each target to look for neighboring stars (see Figure~\ref{fig:k2lc}). 
We use a flag of ``Y'' for yes, ``M'' for maybe, and ``N'' for no to indicate whether the target and a neighbor have blended PSFs on the \textit{K2} chip.
Stars flagged as ``Y'' are labeled candidate binaries; we find 159 such targets, or 23\% of stars with \textit{K2} $P_{rot}$. 
 
To determine the likelihood that these are chance alignments, we offset the cluster positions by 15\degree\ in both RA and Dec and search for neighbors in the SDSS DR12.
We restrict this search to objects with $g\le22$~mag, the SDSS 95\% completeness limit.
We find an SDSS object within 10\arcsec\ (20\arcsec) of 8\% (13\%) of these offset positions. 
This suggests that at least 10\% of Praesepe members have a very wide but bound companion, with separations on the order of $10^3$--$10^4$~AU at Praesepe's distance \citep[181.5$\pm$6~pc;][]{vanleeuwen2009}.
The other neighboring stars are likely background stars that could still contribute flux to the \textit{K2} light curve.
Lacking the observations to confirm which neighboring stars are actually bound companions,
we consider all these stars to be candidate binaries in our analysis.

\begin{figure}[ht]
\centerline{\includegraphics[width=\columnwidth]{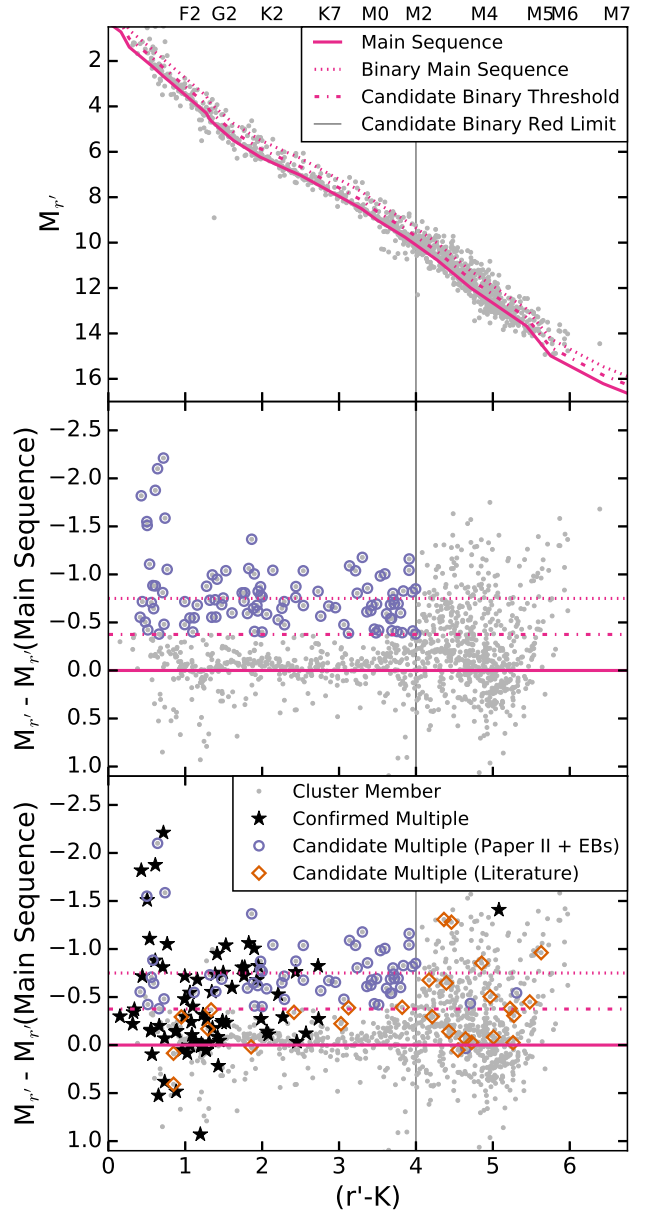}}
\caption{
{\it Top} --- CMD of Praesepe. The solid line is the single-star MS, identified using the spectal energy distributions assembled by \citet{adam2007},
and the dotted line the binary MS. 
We label any star above the dot-dashed line halfway between these two sequences as a candidate binary \citep[as in][]{hodgkin1999}. Stars with $(r'-K)\ \gapprox\ 4$ have a wider range of magnitudes at a given color, and do not show an obvious MS, so we do not identify candidate photometric binaries in this color range.
{\it Middle} --- Residuals between each star's $M_{r'}$ and the model MS magnitude. Photometrically identified candidate binaries are shown as purple circles. This method is primarily sensitive to $\approx$equal-mass binaries.
{\it Bottom} --- Same as above, with confirmed and candidate binaries from the literature shown as black stars and open orange diamonds, respectively. Confirmed binaries can be found at all distances from the MS, illustrating the limitations of this approach to binary identification.}
\label{fig:binarycmd}
\end{figure}

\subsection{Photometric Identification}\label{photbin}
As in previous work, we identify candidate unresolved binaries that are overluminous for their color (see Figure~\ref{fig:binarycmd}). We identify a binary main sequence (MS) offset by 0.75~mag for a given color from that of single stars \citep[as in][]{steele1995}. We then label as candidate binary systems stars with $(r'-K)<4$ that lie above the midpoint between the single-star and binary MSs \citep{hodgkin1999}. This method is biased towards binaries with equal masses, so that we are certainly missing candidate binaries with lower mass ratios. Indeed, confirmed multiples appear at all distances from the putatively single MS (as shown in Figure~\ref{fig:binarycmd}; also see figure 3 in Paper III for a similar analysis in the better-surveyed Hyades). Further observations and analysis are required to confirm the binary status of all cluster members.

We only apply this method to stars with $ (r'-K) <4 $ because the single-star MS is less apparent for stars redder than this value. 
The observed spread in magnitudes could be due to binary systems at a variety of mass ratios, or to increased photometric uncertainties for these faint red stars. Identifying candidate binaries in this regime requires more information than just photometry.

\subsection{Literature Identifications}\label{litbin}

Surveys for multiple systems in Praesepe have been undertaken using lunar occulations \citep{peterson1984, peterson1989}, spectroscopy \citep{mermilliod1990, bolte1991, abt1999, mermilliod1999,halbwachs2003}, speckle imaging \citep{mason1993, patience2002}, adaptive optics imaging \citep{bouvier2001}, and time-domain photometry \citep[e.g.,][]{pepper08}. Spectroscopic binaries in Praesepe have also been identified through larger radial velocity (RV) surveys \citep{pourbaix2004, mermilliod2009}. Several of these surveys also note RV-variable or candidate binary systems. \citet{bolte1991} and \citet{hodgkin1999} identify candidate binary systems by their position above the cluster main sequence (similar to our method above).

Three planets have been detected from RV observations of two Praesepe members \citep{quinn2012, malavolta2016}, including one hot Jupiter in each system.
One confirmed and eight candidate transiting planets have also been discovered from the {\it K2} data for the cluster
\citep{pope2016,barros2016,libralato2016,obermeier2016,mann2016}.


\floattable
\begin{deluxetable}{llcccccccccccll}
[t]\tablewidth{0pt}\tabletypesize{\scriptsize}
\tablecaption{$P_{rot}$ measurements for Praesepe stars targeted in K2 \label{tab:k2}}
\tablehead{\colhead{Name\tablenotemark{a}} & \colhead{EPIC} 
& \colhead{$P_{rot,1}$} & \colhead{Power$_{1}$} & \colhead{$Q_{1}$} 
& \colhead{Clean} & \colhead{Threshold} 
& \colhead{$P_{rot,2}$} & \colhead{Power$_{2}$} & \colhead{$Q_{2}$} 
& \colhead{Multi} & \colhead{Spot} & \colhead{Blend} & \colhead{Bin.}
& \colhead{Ampl.(mag)} 
}
\startdata
JC201 & 211930461 & 14.59 & 0.83910 & 0 & Y & 0.00816 & \nodata & \nodata & \nodata & N & Y & Y & Conf & 0.00934 \\
\nodata & 212094548 & 6.60 & 0.00890 & 1 & N & 0.00521 & \nodata & \nodata & \nodata & N & N & N & \nodata & 0.04953  \\
\nodata & 211907293 & \nodata & \nodata & 2 & \nodata & \nodata & \nodata & \nodata & \nodata & \nodata & \nodata & \nodata & 0 & \nodata  \\
KW222 & 211988287 & 3.29 & 0.20730 & 0 & N & 0.00861 & \nodata & \nodata & \nodata & M & Y & N & \nodata & 0.00281 \\
KW238 & 211971871 & 2.96 & 0.66740 & 0 & Y & 0.00747 & \nodata & \nodata & \nodata & N & Y & N & \nodata & 0.01633  \\
KW239 & 211992776 & 1.18 & 0.30260 & 0 & Y & 0.00791 & \nodata & \nodata & \nodata & M & Y & N & \nodata & 0.00109   \\
KW282 & 211990908 & 2.56 & 0.25900 & 0 & Y & 0.00802 & \nodata & \nodata & \nodata & Y & Y & Y & Conf & 0.00497 \\
AD 2305 & 212100611 & 1.34 & 0.44670 & 0 & Y & 0.00776 & 1.8001 & 0.18730 & 0 & Y & N & M & \nodata & 0.03630  \\
AD 2482 & 211795467 & 15.49 & 0.23690 & 0 & Y & 0.00794 & \nodata & \nodata & \nodata & N & Y & N & \nodata & 0.01088  \\
JS352 & 211913532 & 16.33 & 0.09270 & 0 & Y & 0.00723 & 2.8027 & 0.01060 & 0 & Y & Y & N & Cand & 0.01570  \\
\enddata
\tablecomments{This table is available in its entirety in a machine-readable form in the online journal. A portion is shown here for guidance regarding its form and content.}
\tablenotetext{a}{Literature name given in \citet{adam2007}. All are standard SIMBAD identifiers, except AD\#\#\#\#, which correspond to stars in \citet{adams2002}.}
\end{deluxetable}

\subsection{Binaries Identified from K2 Data}

No eclipsing binaries in Praesepe have been published from the {\it K2} data so far, but we identify four likely eclipsing binaries and two single-transit events by eye; see Appendix \ref{eclipses} for details. 
One of these candidate eclipsing binaries was previously identified from PTF data, and has been confirmed with RVs (Kraus et al.\ in prep.).
We consider the other three eclipsing binaries to be candidate binaries until we can confirm that the eclipses are not from a background system.

We also consider stars with multiple periods visible in the {\it K2} light curve to be candidate binaries if the two peaks are separated by at least 20\% of the primary period. In other words, if
\begin{equation}
\frac{|P_{rot,1} - P_{rot,2}|}{P_{rot,1}} > 0.2,
\end{equation}
we consider the target to be a candidate binary. 
This threshold is based on the maximum period separation for differentially rotating spot groups on the Sun \citep[c.f.~][]{rebull2016-1}. 
Fifty-eight \textit{K2} targets have a second period detected in their periodogram, and nine more have a second period identified by eye only, giving 67 (10\%) multiperiodic targets out of the 677 {\it K2} targets with measured $P_{rot}$. 

In total, we find 82 confirmed binaries or triples, 92 candidate systems from our photometric analysis and the literature, and 170 additional candidate systems identified from our {\it K2} analysis. 
Table \ref{tab:bin} lists the binary members and their relevant properties, and they are also flagged in Table~\ref{tab:k2}.
Aside from the M-dwarf eclipsing binary noted above, however, confirmed binaries in Praesepe are only found above $0.72~\Msun$, which limits our ability to analyze the impact of binaries on rotation and activity in low-mass Praesepe members.

\begin{figure}[t]
\centerline{\includegraphics[width=\columnwidth]{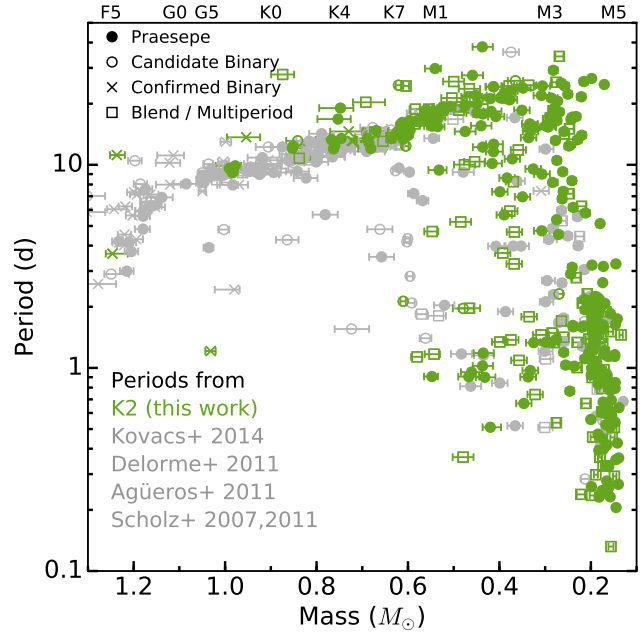}}
\caption{Praesepe mass-period plane showing literature (grey) and previously unpublished high-quality {\it K2} (green) $P_{rot}$ for stars with $P_{mem}>70$\%. We also mark confirmed and candidate binaries: crosses indicate confirmed binaries, open circles indicate photometric or spectroscopic candidate binaries, and open squares indicate {\it K2} targets with a blended neighbor or a second period in the light curve. 
Approximate spectral types are indicated along the top.}
\label{fig:periodmass}
\end{figure}

\section{Results and Discussion}\label{res}
We measure $P_{rot}$ for 677 Praesepe members with \textit{K2}, or 85\% of the 794 Praesepe members with $P_{mem}>50$\% and \textit{K2} light curves. Of these, 471 are new measurements, and 398 (84\%) of these are considered high quality, meaning the periodogram is clean and our by-eye quality flag Q~$=0$ (see Section \ref{validation}).
This sample excludes 94 $P_{rot}$ detections ($12\%$ of the original sample) that we flag as spurious and remove, along with 23 stars ($3\%$) whose periodograms lack significant  peaks. 
The cluster's updated mass-period distribution is shown in Figure~\ref{fig:periodmass}. 
In addition to confirmed and candidate photometric or spectroscopic binaries, we also indicate cases where two or more stars may be contributing to the {\it K2} light curves: open squares are targets with blended neighbors or that show multiple periodic signals.


\subsection{Consistency of $P_{rot}$ Measured from Different Surveys}\label{consistency}

There are 207 Praesepe members with $P_{rot}$ measured both from {\it K2} data and from at least one ground-based survey.
Another 51 members have $P_{rot}$ measured by multiple ground-based surveys, but not by {\it K2}.  
The 43 stars with $P_{rot}$ from at least two studies that differ by $>$10\% are listed in Table \ref{tab:protdiff}. 
Overall, the agreement between \textit{K2} and literature $P_{rot}$ measurements is excellent: half of our \textit{K2} measurements are consistent with previous measurements to within 2\%, and $>$75\% are consistent to within 5\%.

Discrepant measurements are typically $\frac{1}{2}\times$ or $2\times$ harmonics of each other. All but two stars with discrepant $P_{rot}$ show evidence of evolving spot configurations: either a double-dip light-curve structure or a varying amplitude of modulation over the course of the campaign. This signature is usually better resolved in the {\it K2} light curves, allowing us to measure the correct period even if it is not the highest periodogram peak. 

Additionally, four stars with discrepant $P_{rot}$ values show evidence for two periods in the light curve.
The PTF and \textit{K2} periods for EPIC 211937872 and EPIC 211971354 are $\approx$1~d apart; both \textit{K2} light curves show a second $\approx$1~d period superimposed on the primary period, in addition to evidence of spot evolution.
Two periods are detected in the \textit{K2} data for EPIC 212013132: $P_{rot,1}=2.13$~d, half of the SWASP $P_{rot}=4.27$~d, and $P_{rot,2}=12.32$~d, consistent with the \citet{kovacs2014} $P_{rot}=12.78$~d.
Finally, two periods are also detected in the \textit{K2} light curve of EPIC 211734093: $P_{rot,1}=18.22$~d and $P_{rot,2}=7.74$~d. The latter of these is half of the \citet{kovacs2014} $P_{rot}=15.87$~d. 
In all four cases, the second period in the light curve, possibly with additional spot evolution effects, accounts for the discrepancy between our \textit{K2} $P_{rot}$ values and those in the literature. 


In three cases, the $P_{rot}$ measured by \citet{scholz2007} and \citet{scholz2011} is potentially a 1-d alias of the {\it K2} period.
\citet{scholz2007} surveyed Praesepe over three observing runs lasting three to five nights each, and \citet{scholz2011} surveyed the cluster again for nine nights. Measurements over such short baselines are more prone to aliasing, particularly when the periods are so close to 0.5 or 1 d.

We find no strong evidence that the Praesepe stars with literature $P_{rot}$ have larger photometric amplitudes (Figure~\ref{fig:amp}), which has often been invoked to explain low $P_{rot}$ yields from ground-based surveys. 
The only partial exception to this are the PTF data: in Paper I, we could only measure $P_{rot}$ for stars with amplitudes $\gapprox$0.02 mag ($\gapprox$1\%) for $K_p>16$. 
Aside from this handful of PTF stars, small photometric amplitude---i.e., less contrast between starspots and the stellar photosphere---does not explain the incompleteness of ground-based surveys.

\begin{figure}[t]
\centerline{\includegraphics[width=\columnwidth]{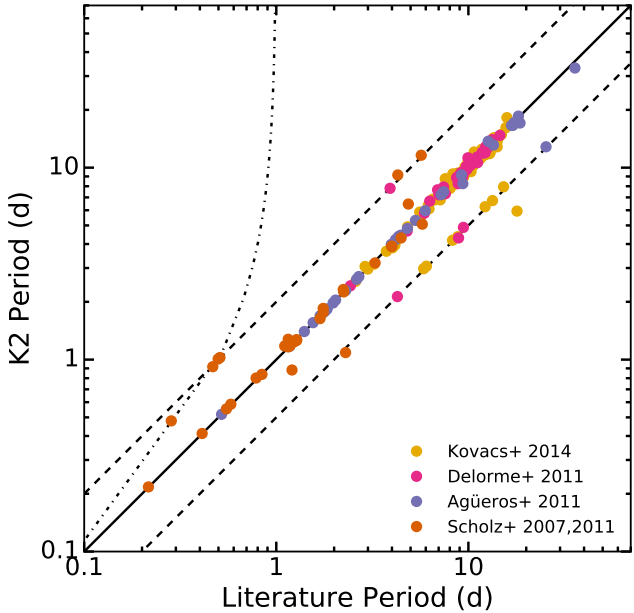}}
\caption{$P_{rot}$ from {\it K2} compared to literature $P_{rot}$ for the same stars. Colors are the same as in Figure~\ref{fig:periodmass_old}. Our new $P_{rot}$ are generally consistent with literature $P_{rot}$, except for a handful of cases where the older $P_{rot}$ is a harmonic (dashed lines) or 1-d alias (dot-dashed line) of the {\it K2} $P_{rot}$.}
\label{fig:compk2}
\end{figure}

\begin{figure}[t]
\centerline{\includegraphics[width=\columnwidth]{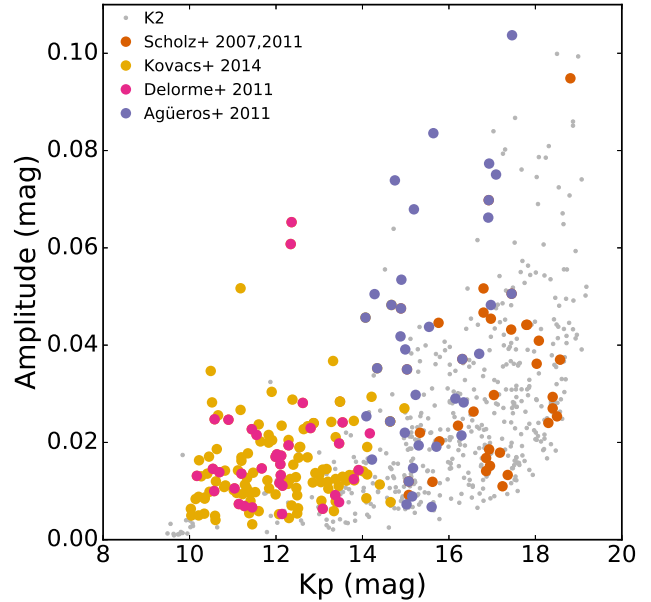}}
\caption{Amplitude of the \textit{K2} light curve as a function of $K_p$; colors indicate stars with literature $P_{rot}$ and are the same as in Figure \ref{fig:periodmass_old}. We find no strong evidence that stars with $P_{rot}$ measured by ground-based surveys have higher amplitudes. Two targets with amplitudes $>$0.2~mag are not shown for clarity. }
\label{fig:amp}
\end{figure}

Overall, 86\% of Praesepe {\it K2} targets have detectable $P_{rot}$, suggesting that non-detections in ground-based surveys are due primarily to limitations of those surveys rather than to inclination effects or spot coverage.
Our Praesepe and Hyades {\it K2} $P_{rot}$ are nearly all consistent with previous measurements (Figure~\ref{fig:compk2} and Paper III).
We conclude that ground-based $P_{rot}$ measurements are reasonably reliable, and that these surveys are merely limited by trade-offs between photometric precision, cadence, baseline, and number of targets; interruptions due to daylight and weather; and variable spot patterns on the stars themselves.
Further comparisons of the \textit{K2} data with ground-based light curves and survey techniques are needed to determine why previous surveys did not detect the rotators with new $P_{rot}$ measured here. 

Nonetheless, the overall agreement between the \textit{K2} measurements and those of previous surveys indicates that our $P_{rot}$ measurement procedures provide accurate results. It also bodes well for future ground-based surveys: while \textit{K2}'s superior precision allows us to resolve detailed light-curve features, it appears that in general, ground-based surveys produce valid and reproducible $P_{rot}$ measurements.

\subsection{Binaries in the Mass-Period Plane}\label{bindisc}

In Paper III, we found that nearly all the rapid rotators in the Hyades with $M_*\ \gapprox\ 0.3$~\Msun\ were confirmed or candidate binary systems. Of the three remaining rapid rotators, none had been surveyed for companions. 
The Hyades as a whole has been extensively surveyed for companions: $>$30\% of all Hyads are confirmed binaries, including $\approx$45\% of Hyads with measured $P_{rot}$.
This suggested that all single stars with $M_*\ \gapprox\ 0.3$~\Msun\ have converged onto the slow-rotator sequence by $\approx$650~Myr.

\begin{figure}[t]
\centerline{\includegraphics[width=\columnwidth]{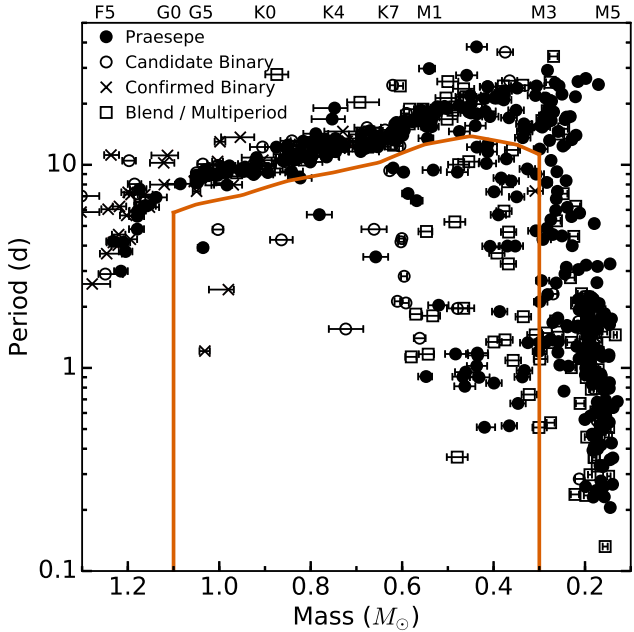}}
\caption{Mass-period plane with the region that defines $M_*>0.3$~\Msun\ rapid rotators outlined in orange. Half of the rapid rotators are confirmed or candidate binaries.
}
\label{fig:rapid}
\end{figure}

For Praesepe stars, we define the cutoff between the slow-rotator sequence and more rapid rotators by computing the 75$^{\rm th}$ percentile of periods for stars with $1.1\ \gapprox\ M_*\ \gapprox\ 0.3$~\Msun, and then lowering this threshold by 30\%. 
This produces the orange line shown in Figure \ref{fig:rapid}. 
We find that half of all rapidly rotating Praesepe stars are confirmed or candidate binaries. 

Despite the far more extensive $P_{rot}$ catalog in Praesepe relative to the Hyades, however, we are currently unable to confirm our result from Paper III because Praesepe lacks a similarly rich binary catalog. 
Only 7\% of all cluster members are confirmed binaries, and (with the exception of one eclipsing M-dwarf binary) these are restricted to $M_*\ \gapprox\ 0.72$~\Msun. 
Our identification of candidate systems is also likely incomplete. 
Confident analysis of the impact of binaries on the mass-period plane requires additional binary searches in Praesepe.

Many stars on the slow-rotator sequence are also candidate binaries. 
This might suggest that companions have minimal impact on angular-momentum evolution. 
It could also indicate that different subsets of binaries undergo different rotational evolution.

The rapidly and slowly rotating binaries likely have different separation distributions, due to the impact of disk disruption on their initial angular-momentum content. 
Single stars experience braking due to their protoplanetary disks \citep{rebull2004}.
Binaries wider than 40~AU are unlikely to disrupt each others' protoplanetary disks \citep{cieza2009,kraus2016} and are far too wide to be affected by tides---these systems will therefore evolve as (two) single stars. 
Binaries closer than 40~AU, on the other hand, are far more likely to have disrupted disks, which would allow the component stars to spin up without the losing angular momentum to their disks. 
These systems will arrive on the MS spinning more rapidly, and eventually spin down to converge with single stars. 
We expect that future studies of Praesepe will find that binaries with slowly-rotating components are wider than 40~AU ($\approx$0.2$\arcsec$ at $\approx$180~pc), while the rapidly rotating stars have companions at closer separations.

\subsection{Comparison with Models of Rotation Evolution}\label{gyro}

In Paper III, we found that the \citet{reiners2012} and \citet{matt2015} models for angular-momentum evolution predicted faster rotation than observed for 0.9$-$0.3~\Msun\ stars.
However, this comparison was limited by the number of Hyads with $P_{rot}$. 
We therefore compare our far richer Praesepe sample with the models of \citet{matt2015} and \citet{brown2014}, which were generously provided by these authors (S.~Matt, private communication, 2015; T.~Brown, private communication, 2017).

\subsubsection{Matt et al.\ (2015)}

\citet{matt2015} derive a model for the angular-momentum evolution of a rotating solid sphere due to magnetic braking. 
These authors' initial conditions approximate the distribution of $P_{rot}$ observed for 2$-$5-Myr-old stars, but are not drawn directly from observations. 
\citet{matt2015} allow the stellar radius to evolve according to model evolutionary tracks. 
Their prescription for the angular momentum lost via stellar winds is based on the \citet{kawaler1988} and \citet{matt2012} solar-wind models, 
and the angular-momentum loss scales with stellar mass and radius. 
\citet{matt2015} also use explicitly different spin-down rates for stars in the saturated and unsaturated regime. 

\begin{figure}[t]
\centerline{\includegraphics[width=\columnwidth]{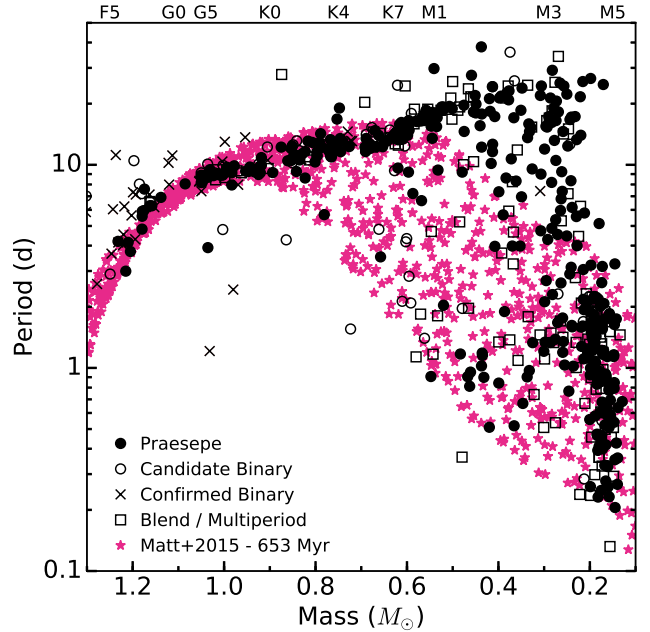}}
\caption{Comparison of $P_{rot}$ in Praesepe (black; symbols are as in Figure \ref{fig:periodmass}) with the $P_{rot}$ distribution predicted by \citet{matt2015} at 653~Myr (purple stars).
Only literature $P_{rot}$ and clean, Q~$=0$ {\it K2} detections are shown. 
The model matches the slow-rotator sequence for single $>$0.8~\Msun\ stars, but fails to predict that the majority of 0.6$-$0.3~\Msun\ stars are slowly rotating.
}
\label{fig:matt}
\end{figure}

\begin{figure}[t]
\centerline{\includegraphics[width=\columnwidth]{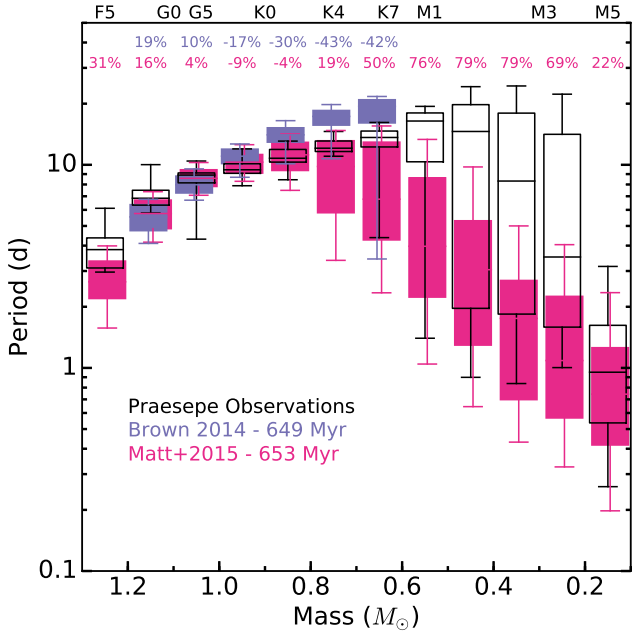}}
\caption{
A box-and-whiskers plot of the Praesepe mass-period plane (black) compared to predictions by \citet[][purple]{brown2014} and \citet[][pink]{matt2015}. 
The boxes extend to the 25$^{\rm th}$ and 75$^{\rm th}$ percentiles, and the whiskers extend to the 5$^{\rm th}$ and 95$^{\rm th}$ percentiles. 
The percentage differences between the data and model medians are printed across the top of the plot.
The \citet{matt2015} models fail to predict the slow rotation periods for the majority of $<$0.6~\Msun\ stars. 
}
\label{fig:boxplot}
\end{figure}

The \citet{matt2015} model accurately predicts the mass dependence of the slow-rotator sequence for Hyades and Praesepe stars with $M_*\ \gapprox\ 0.8\ \Msun$, with the exception of a handful of binary stars (see Figures \ref{fig:matt}--\ref{fig:boxplot}). 
This indicates that, as in our comparison to the Hyades alone, the stellar-wind prescription used by \citet{matt2015} is correct for solar-type stars. 

\begin{figure}[t]
 \centerline{\includegraphics[width=0.96\columnwidth]{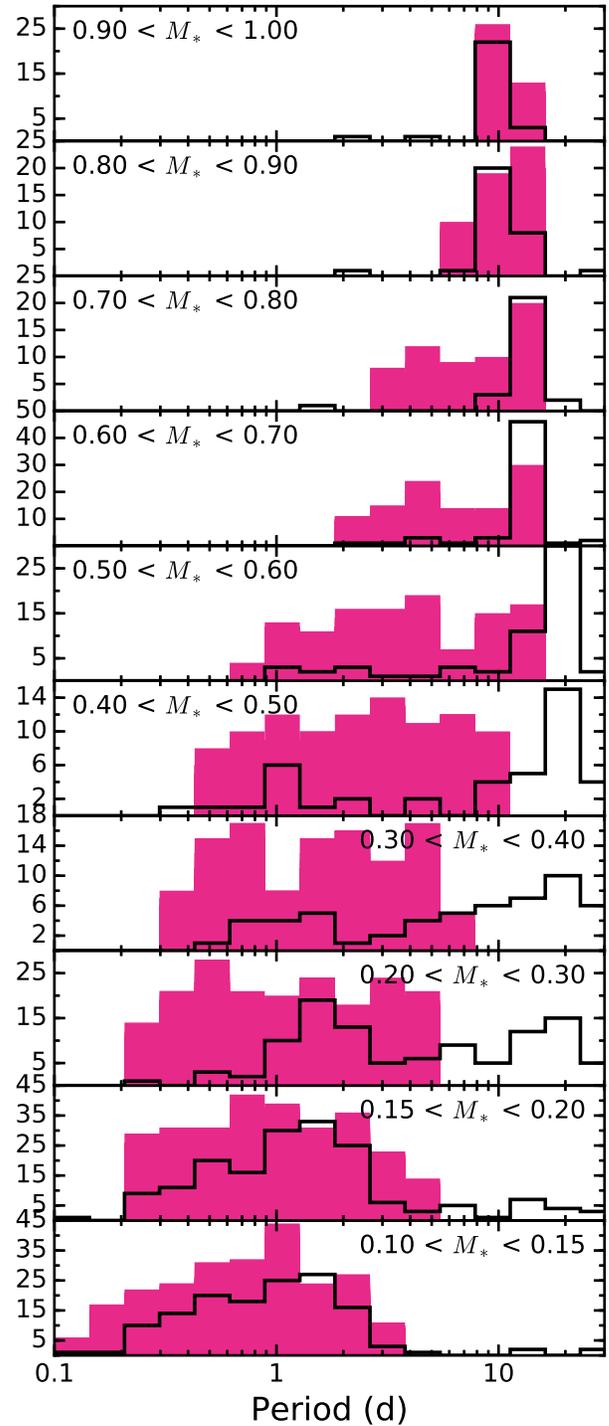}}
\caption{ 
Praesepe $P_{rot}$ distribution  (solid line) compared to that predicted by \citet[][shaded]{matt2015} for different mass bins. 
The mass bins are smaller at the lowest masses and larger for solar-type stars.
The histograms represent 200 randomly drawn sets of modeled points; each set contains the same number of stars observed in that mass bin. These random subsets are plotted with transparency, so that the model histograms are darker when they are more frequently produced at that height.  
The model accurately tracks the slow-rotator sequence for $>$0.6~\Msun\ stars, but fails to predict the majority of slowly rotating M dwarfs. 
}
\label{fig:matt_hist}
\vspace{-0.4cm}
\end{figure}

The lower envelope of $P_{rot}$ predicted by \citet{matt2015} approximates that observed in Praesepe, although the distribution of rapidly rotating stars with $M_*~\lapprox~0.8~\Msun$ is much more sparse than predicted by the model. 
Using the division between the slow sequence and faster rotators defined in Section~\ref{bindisc}, we observe that 26\% of stars with masses 0.3$-$0.8~\Msun\ are rapidly rotating, relative to 77\% of model stars. 
In Figure~\ref{fig:matt_hist}, we have binned the model and data $P_{rot}$ distributions by mass to allow for easier comparisons of the period distribution. 
Below $\approx$0.8~\Msun, the \citet{matt2015} model predicts a broader distribution of periods than is observed, while the observed $P_{rot}$ are more concentrated at slow periods with a tail of fast rotators. 
This suggests that although the \citet{matt2015} model may work for some 0.8$-$0.3~\Msun\ stars, it fails to predict the efficiency with which $<$50\% of stars in this mass range spin down. 

The most obvious discrepancy between the \citet{matt2015} models and our data occurs for slowly rotating early M stars with masses $\approx$0.6$-$0.3~\Msun, as was previously noted in \citet{matt2015} and Paper III.
In our observations, more than half of the 0.6$-$0.3~\Msun\ stars have converged to the slow-rotator sequence, which extends fairly smoothly from $\approx$1$-$0.3~\Msun\ (see Figure~\ref{fig:boxplot}), and more than half of the remaining rapid rotators are binaries (Figure~\ref{fig:matt}).

\begin{figure}[t]
\centerline{\includegraphics[width=\columnwidth]{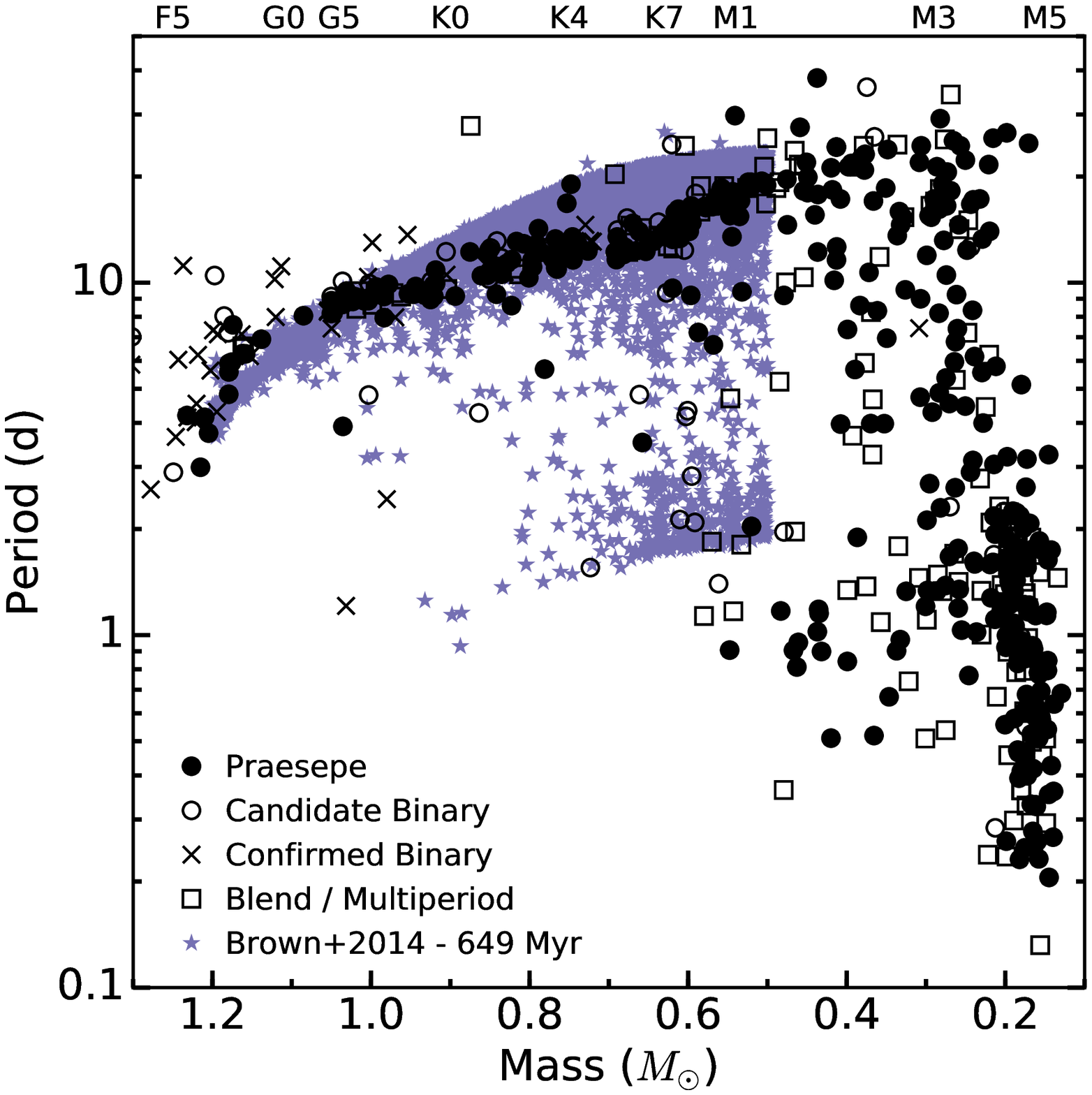}}
\caption{Comparison of $P_{rot}$ in Praesepe (black; symbols are as in Figure~\ref{fig:periodmass}) with the $P_{rot}$ distribution predicted by \citet{brown2014} at 649~Myr (purple stars). 
Only literature $P_{rot}$ and clean, Q~$=0$ {\it K2} detections  are shown. 
This model only covers $\approx$1.2$-$0.5~\Msun, and predicts a bimodal $P_{rot}$ distribution. However, we do not observe a strongly bimodal distribution in Praesepe, and the model fails to predict the rapidly rotating Praesepe stars with $M_*\approx0.6$~\Msun\ and $P_{rot}\approx1$~d.
}
\label{fig:brown}
\end{figure}

By contrast, the model predicts an end to the slow-rotator sequence around 0.6~\Msun, with the slowest rotators at lower masses being significantly faster than the slow rotators observed in our data. 
The median $P_{rot}$ we observe for 0.6$-$0.3~\Msun\ stars is $>$75\% slower than predicted (Figure~\ref{fig:boxplot}).
Furthermore, $>$60\% of stars in this mass range rotate more slowly than the maximum $P_{rot}$ predicted for their mass (Figure~\ref{fig:matt_hist}).
It appears that real early M dwarfs brake far more efficiently than predicted by \citet{matt2015}. 

This discrepancy suggests that most M dwarfs undergo enhanced angular-momentum loss relative to their higher mass counterparts.
This could be due to a change in the structure of the magnetic field , i.e., a larger, less complex field with more open field lines near the star's equator that would allow the star to more efficiently shed angular momentum \citep[i.e.,][]{donati2011,garraffo2015}.
It could also indicate a departure from solid-body rotation, which is assumed by \citet{matt2015}, for early M dwarfs. 
We could be observing an effect of the deepening convective zone for M dwarfs, through a change in the moment of inertia or in the dynamo as the radiative core shrinks with decreasing mass. 

Despite the failure of the \citet{matt2015} model to predict the observed behavior of early M dwarfs, this model does reasonably well in reproducing the distribution of rapidly rotating, fully convective 0.1$-$0.2~\Msun\ stars. This suggests that whatever is to blame for the discrepancy with observed early M dwarfs, the physical assumptions of \citet{matt2015} do apply to fully convective stars.

\begin{figure}[t]
\centerline{\includegraphics[width=0.96\columnwidth]{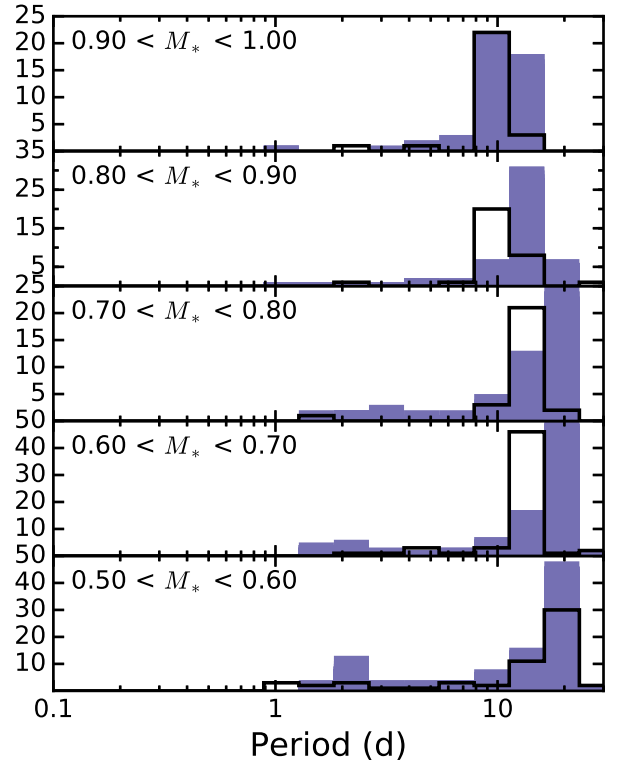}}
\caption{
Praesepe $P_{rot}$ distribution (solid line) compared to that predicted by \citet[][shaded]{brown2014} for different mass bins. Here again the mass bins are not even, and this model does not extend below 0.5~\Msun.
The histograms represent 200 randomly drawn sets of the modeled points; each set contains the same number of stars observed in that mass bin. 
This model correctly predicts that most stars will have converged onto the slow-rotator sequence by this age, but fails to predict the evolution of faster rotators. 
}
\label{fig:brown_hist}
\end{figure}

\subsubsection{Brown\ (2014): The Metastable Dynamo Model}

\citet{brown2014} derives an empirical model for the generation and evolution of the fast and slow rotator sequences, called the Metastable Dynamo Model (MDM). 
He models all stars as solid bodies that are born with weak coupling between their dynamos and stellar winds, leading to minimal spin-down.
The stars then spontaneously and permanently switch into a strong coupling mode where they spin down rapidly. 
\citet{brown2014} does not employ a critical $P_{rot}$ for this switch---it is purely stochastic, with a mass-dependent probability of switching by a given stellar age. 
Taking as its starting point the distribution of periods in the 13-Myr-old cluster h Per, the MDM generates a bimodal distribution at older ages: a fast sequence and a slow sequence separated by a gap, similar to the distribution observed by \citet{barnes2003}. 

The \citet{brown2014} model approximately reproduces the overall morphology of the mass-period plane in Praesepe: there is a clear sequence of slowly rotating stars with some faster rotators. A more careful comparison, however, indicates that the  model and data are discrepant.
Specifically, the bimodality is not obvious in the data for Praesepe $>$0.5~\Msun\ stars, which is the mass regime covered by the MDM (Figures~\ref{fig:brown}-\ref{fig:brown_hist}). 
The rapidly rotating Praesepe stars in this mass range are not strongly concentrated at any particular $P_{rot}$, nor is there an obvious gap at intermediate $P_{rot}$.
Furthermore, using the division between slow and fast stars defined in Section~\ref{bindisc}, we find that 15\% of observed 0.9$-$0.5~\Msun\ Praesepe stars are rapidly rotating, compared to only 7\% in the model.

We do observe stronger bimodality for 0.25$-$0.5~\Msun\ stars, below the mass range modeled by the \cite{brown2014}. 
The observed morphology does not match the predictions for $>$0.5~\Msun\ stars, however: the rapid rotators extend to $P_{rot}\approx\frac{1}{2}$~d and show a wider range of fast $P_{rot}$, in contrast to the clear lower limit of $\approx$1.5~d in the current model. 
Our observations of early M stars in Praesepe therefore support the \citet{brown2014} model's prediction of bimodality in the mass-period morphology, but adjustments are needed to extend the MDM to this mass range.

Finally, the predicted locations of the fastest and slowest rotators at a given mass do not match the observations. The slow-rotator sequence is too slow, while a handful of early M dwarfs with $0.5\ \lapprox\ M_*\ \lapprox\ 0.6$~\Msun\ rotate faster than predicted by the MDM. 
The offset of the slow sequence is visible in figure 6 of \citet{brown2014}, who points out that more complicated physics is likely needed to explain the exact evolution of slow rotators. The too-fast rotators are not obvious in that figure, however, due to the use of a linear $P_{rot}$ axis and the inclusion of only a few dozen $P_{rot}$ from \citet{delorme2011} and WEBDA, compared to the hundreds of $P_{rot}$ included here.

\section{Conclusions}\label{concl}

We analyze {\it K2} light curves for 794 members of the Praesepe open cluster, and present $P_{rot}$ for 677 {\it K2} targets. 
Of these, 471 are new measurements, bringing the total number of $P_{rot}$ measurements for Praesepe members to 732.

We find that half of the rapidly rotating stars with $M_*\ \gapprox\ 0.3\ \Msun$ are confirmed or candidate binary systems. The remaining $\gapprox0.3$~\Msun\ fast rotators are not confirmed single stars, as they have not been searched for binary companions.  We previously found that all rapidly rotating $\gapprox0.3$~\Msun\ Hyads are binaries \citep{douglas2016}, but we require deeper binary searches in Praesepe to confirm whether binaries in the two co-eval clusters have different $P_{rot}$ distributions.

We also compare the $P_{rot}$ distribution in Praesepe to that predicted by \citet{matt2015} and \citet{brown2014} for $\approx$650~Myr-old stars. 
We find that \citet{matt2015} correctly predict the slow rotator sequence for $>$0.8~\Msun\ stars, but that $>$60\% of $\approx$0.6$-$0.3~\Msun\ stars are rotating more slowly than predicted. 
This suggests that a change in braking efficiency occurs for early M dwarfs, causing them to spin down more quickly than predicted using a scaled solar-wind model. 
We do not observe a clear bimodality in $P_{rot}$ for Praesepe stars with $M_*>0.5$~\Msun, in contrast with the \citet{brown2014} model predictions. We do observe stronger bimodality for 0.25$-$0.5~\Msun\ stars, but adjustments will likely be needed to extend the model to this mass range.

Binaries likely impact our comparison with these models, which assume that stars evolve in isolation. 
This should work well for actual single stars, of course, as well as for wider binaries that never interact, but not for closer binaries, many of which have yet to be identified in these open clusters. 
If most or all rapidly rotating stars are binaries, and particularly if their rapid rotation is due to increased initial angular-momentum content, then it is unsurprising that models struggle to replicate simultaneously the distributions of slow and rapid rotators. Theorists may be attempting to match a population of stars reflecting a set of initial conditions that do not match their assumptions. 
Confirmed single stars will be better calibrators for these models, and binary surveys of Praesepe will be crucial for obtaining a proper benchmark sample.

\acknowledgments

M.A.A.~acknowledges support provided by the NSF through grants AST-1255419 and AST-1517367. 
K.R.C.~acknowledges support provided by the NSF through grant AST-1449476. 
We thank Tim Brown and Sean Matt for sharing their models with us, as well as Rob Jeffries, Sean Matt, and Marc Pinsonneault for useful discussions and advice. We also thank the anonymous referee for comments which improved the manuscript.

We have made use of the {\it Astropy} package \citep{astropy}, 
the {\it Astropy}-affiliated {\it Astroquery} package \citep{ginsburg2013},  
the {\it AstroML} package \citep{vanderplas2012,ivezic2013}, 
the {\it pywcsgrid2} package developed by J.~Lee, \footnote{\url{https://github.com/leejjoon/pywcsgrid2}}
and the {\it K2fov} package created by F.~Mullally, T.~Barclay, and G.~Barentsen for NASA's Kepler/K2 Guest Observer Office \citep{mullally2016}

This research has made use of NASA's Astrophysics Data System Bibliographic Services, the SIMBAD database, operated at CDS, Strasbourg, France, and the VizieR database of astronomical catalogs \citep{Ochsenbein2000}.

This paper includes data collected by the {\it K2} mission. Funding for the {\it K2} mission is provided by the NASA Science Mission directorate. Some of the data presented in this paper were obtained from the Mikulski Archive for Space Telescopes (MAST). STScI is operated by the Association of Universities for Research in Astronomy, Inc., under NASA contract NAS5-26555. Support for MAST for non-{\it HST} data is provided by the NASA Office of Space Science via grant NNX09AF08G and by other grants and contracts.

This research has made use of the NASA/ IPAC Infrared Science Archive, which is operated by the Jet Propulsion Laboratory, California Institute of Technology, under contract with the National Aeronautics and Space Administration. The Two Micron All Sky Survey was a joint project of the University of Massachusetts and IPAC.

The Digitized Sky Survey was produced at the Space Telescope Science Institute under U.S. Government grant NAG W-2166. The images of these surveys are based on photographic data obtained using the Oschin Schmidt Telescope on Palomar Mountain and the UK Schmidt Telescope. The plates were processed into the present compressed digital form with the permission of these institutions.

Funding for SDSS-III has been provided by the Alfred P. Sloan Foundation, the Participating Institutions, the National Science Foundation, and the U.S. Department of Energy Office of Science. The SDSS-III web site is \url{http://www.sdss3.org/}. SDSS-III is managed by the Astrophysical Research Consortium for the Participating Institutions of the SDSS-III Collaboration.

\appendix

\section{Candidate Transiting and Eclipsing Systems}\label{eclipses}

\begin{figure*}[t]
\centerline{\includegraphics[width=\textwidth]{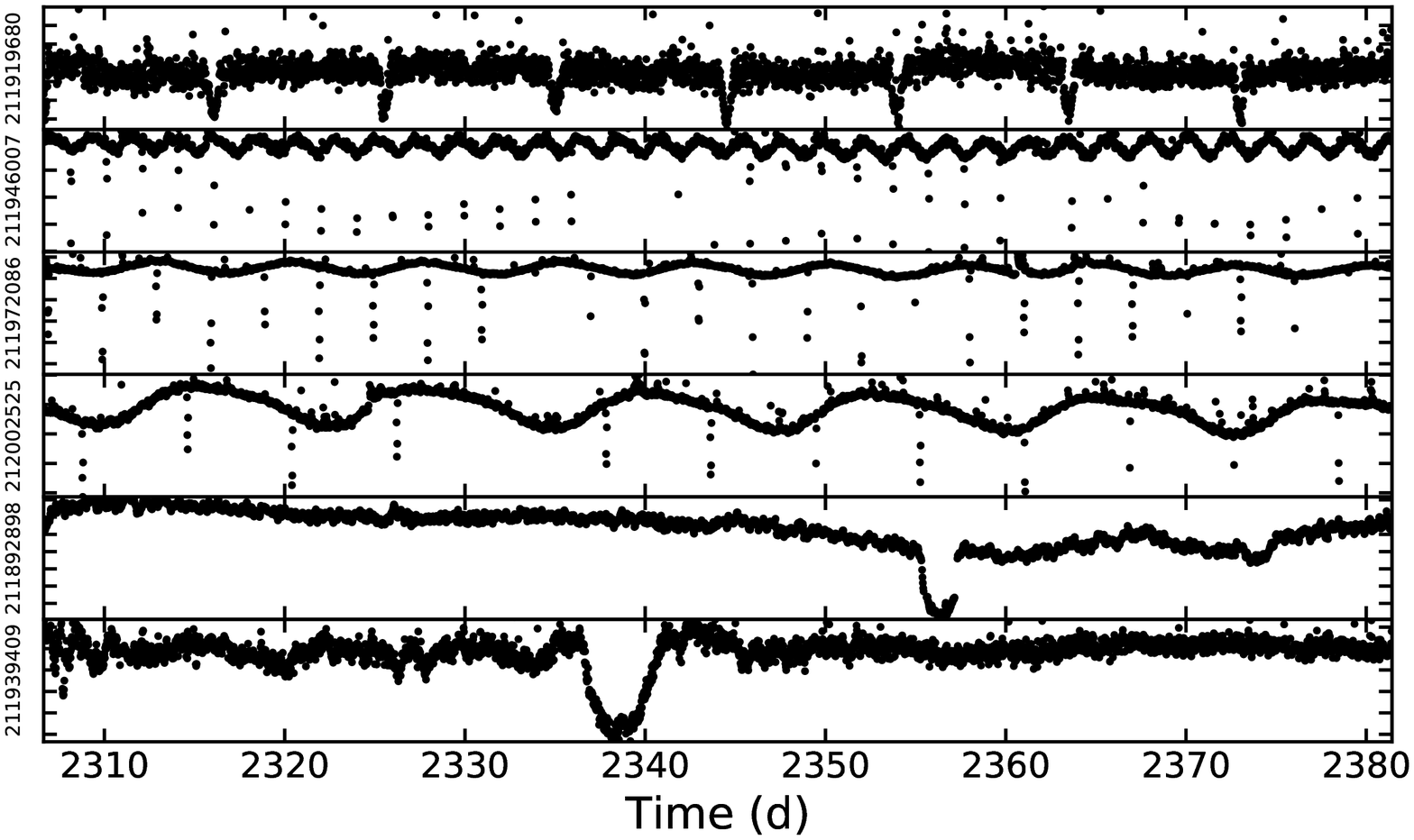}}
\caption{Light curves for seven candidate eclipsing systems identified by eye from Campaign 5 data; EPIC IDs are given on the left axis. The top four systems show signatures of rotation as well as multiple eclipses, and the bottom three show only one eclipse during Campaign 5. EPIC 211972086 (third panel) was preciously detected in PTF data, and has been confirmed with RVs (Kraus et al.\ in prep.)}
\label{fig:eclipses}
\end{figure*}

In our by-eye inspection/validation of the {\it K2} light curves and $P_{rot}$ measurements, we identify six candidate eclipsing systems. We briefly discuss each them here, but with one exception, we make no attempt to confirm them at this time.
The membership probabilities and spectral types noted below are all from \citet{adam2007}.
We present the light curves for these seven objects in Figure~\ref{fig:eclipses}, and the inspection/validation plots for each can be found in the electronic sub-figures for Figure \ref{fig:k2lc} noted below.

$\bullet$ \emph{EPIC 211919680} (2MASS J08440390$+$1901129, HSHJ474, $M_*=0.18~\Msun$, M5, $P_{mem}=96$\%) shows sinusoidal modulation with $P_{rot}=0.31$~d and eclipses every 4.77~d. There is no other star visible nearby in the SDSS $r$-band image. This star has not been previously identified as a binary system.

$\bullet$ \emph{EPIC 211946007} (2MASS J08423944$+$1924520, HSHJ430, $M_*=0.20~\Msun$, M4, $P_{mem}=99$\%) shows sinusoidal modulation with $P_{rot}=2.25$~d, consistent with the $P_{rot}=2.24$~d measured by \citet{scholz2011}, and eclipses every 1.98~d. Three additional stars are visible in the SDSS $r$-band image:  two faint companions near the target star that are blended on the {\it K2} chip, and an additional star just off the edge of the {\it K2} pixel stamp. This star had not been identified as a binary system.

$\bullet$ \emph{EPIC 211972086} (2MASS J08504984$+$1948365,
$M_*=0.31~\Msun$, M3, $P_{mem}=98$\%) was previously identified as a binary from PTF data and has been confirmed via radial velocity (RV) observations; analysis of this system is forthcoming in Kraus et al.~(in prep). The {\it K2} light curve shows sinusoidal modulation with $P_{rot}=7.49$~d, consistent with the PTF $P_{rot}=7.43$~d, as well as eclipses. There is no other star visible nearby in the SDSS $r$-band image. The eclipsing binary (EB) period ($\approx$6~d) is not detected in the periodogram.

$\bullet$ \emph{EPIC 212002525} (2MASS J08394203$+$2017450, M4, $P_{mem}=100$\%) shows sinusoidal modulation with $P_{rot}=12.63$~d; the eclipse period looks slightly shorter than that but is not detected in the periodogram.
There is one star visible in the corner of the {\it K2} pixel stamp.
This star had not been identified as a binary system.

In addition to the above systems, 
EPIC 211929081 and EPIC 211939409
show no sinusoidal modulations but do have a possible single eclipse during Campaign 5. 
These single-eclipse candidates are admittedly more suspect than the four above, as a single drop in flux could be due to any number of instrumental or astrophysical issues. 
The eclipse durations are longer than expected for two main sequence stars eclipsing each other; if these are real astrophysical eclipses, then the eclipse may come from a faint background giant contaminating the light curve, or from a gas giant planet with a large ring system. 
RV data are needed to confirm the cluster membership of these stars and check for companions. 

$\bullet$ \emph{EPIC 211892898} (2MASS J08433463$+$1837199, $M_*=1.06$~\Msun, K4, $P_{mem}=99$\%) has two nearby companions, at least one of which is blended into the {\it K2} PSF of the target. There is also a correlated increase in the white-noise component of the light curve during the eclipse ingress and egress. This star has not been previously identified as a binary system.


$\bullet$ \emph{EPIC 211939409} (2MASS J08512585$+$1918564, 
$M_*=0.36$~Msun, M3, $P_{mem}=96$\%) has a neighboring object in the corner of the {\it K2} pixel stamp. This star has not been previously identified as a binary system.

Only \citet{barros2016} have published EB candidates from Campaign 5, but due to their survey limits these authors did not detect any of the above candidates. 
\citet{barros2016} restricted their analysis to stars with $K_p<15$, which removes the four obvious EB candidates as they all have $K_p>16.5$. \citet{barros2016} also required more than one eclipse or transit for detection, which explains why these authors do not list the two single-eclipse events that we identified by eye.
\citet{barros2016} do not identify any other EB candidates in Praesepe, although as mentioned in Section \ref{binaries} these authors do find one candidate planet in the cluster.

Three other studies have found candidate transiting planets in Praesepe \citep{pope2016,libralato2016,obermeier2016}. These transits are mostly small and were missed when we inspected the light curves.

\setlength{\baselineskip}{0.6\baselineskip}
\bibliography{references}
\setlength{\baselineskip}{1.667\baselineskip}

\end{document}